\pdfoutput=1
\documentclass[a4paper,fleqn,usenatbib]{mnras}
\usepackage[T1]{fontenc}
\usepackage{ae,aecompl}
\usepackage{graphicx}	
\usepackage{amstext}
\usepackage{amsmath}	
\usepackage{amssymb}

\newcommand{\SNII}{SN~II}
\newcommand{\SNIa}{SN~Ia}
\newcommand{\msun}{{\rm M}_{\odot}}

\title[Accretion by stars]{Chemical enrichment of stars due to accretion from the ISM during the Galaxy's assembly}

\author[Shen et al.]
       {Sijing Shen$^{1, 2}$\thanks{E-mail: shens@astro.uio.no}
         Girish Kulkarni$^{2}$,
         Piero Madau$^{3}$
         and Lucio Mayer$^{4}$ \\
         $^{1}$Institute of Theoretical Astrophysics, University of Oslo, Postboks 1029, 0315 Oslo, Norway\\	
         $^{2}$Institute of Astronomy and Kavli
         Institute of Cosmology, University of Cambridge, Madingley Road,
         Cambridge CB3 0HA, UK \\
         $^{3}$Department of Astronomy and Astrophysics, University
         of California, Santa Cruz, 1156 High Street, Santa Cruz, CA
         95064 USA \\
         $^{4}$Institute of Theoretical Physics, University of
         Zurich, Winterthurerstrasse 190, CH-9057 Zurich, Switzerland}
       
\date{Accepted ---. Received ---; in original form ---}

\pubyear{2016}

\begin{document}
\label{firstpage}
\pagerange{\pageref{firstpage}--\pageref{lastpage}}
\maketitle

\begin{abstract}
Using the Eris zoom-in cosmological simulation of assembly of a Milky
Way analog, we study the chemical enrichment of stars due to accretion
of metal-enriched gas from the interstellar medium during the Galaxy's
development.  We consider metal-poor and old stars in the Galactic
halo and bulge through the use of stellar orbits, gas density and
metallicity distributions in Eris.  Assuming spherically symmetric
Bondi-Hoyle accretion, we find that halo and bulge stars accrete
metals at the rate of about $10^{-24}$ M$_\odot$yr$^{-1}$ and
$10^{-22}$ M$_\odot$yr$^{-1}$, respectively, at redshifts $z\lesssim
3$, but this accretion rate increases roughly a hundred-fold to about
$10^{-20}$ M$_\odot$yr$^{-1}$ at higher redshifts due to increased gas
density.  Bulge and halo stars accrete similar amounts of metals at
high redshifts when kinematically distinct bulge and halo have not yet
developed, and both sets of stars encounter a similar metal
distribution in the ISM.  Accretion alone can enrich main-sequence
stars up to $[\mathrm{Fe}/\mathrm{H}]\sim -2$ in extreme cases, with
the median enrichment level due to accretion of about
$[\mathrm{Fe}/\mathrm{H}] \sim -6$ to $-5$.  Because accretion mostly
takes place at high redshifts, it is $\alpha$-enriched to
$[\alpha/\mathrm{Fe}]\sim 0.5$.  We find that accretive metal
enrichment is sufficient to affect the predicted metallicity
distribution function of halo stars at $[\mathrm{Fe}/\mathrm{H}]<-5$.
This can hinder attempts to infer natal chemical environment of
metal-poor stars from their observed enrichment.  Peculiar enrichment
patterns such as those predicted to arise from pair-instability
supernovae could help in disentangling the natal and accreted metal
content of stars.
\end{abstract}


\section{Introduction}

Stars are thought to preserve a chemical record of their natal
environment in their presently observed metal abundance
\citep{2013pss5.book...55F}.  This makes old, low-mass, metal-poor
stars in the Galaxy an important probe of star formation at the
earliest cosmological times \citep{2015ARA&A..53..631F}.  Stars with
masses less than 1~M$_\odot$ have lifetimes longer than 12.7 Gyr
\citep{1989A&A...210..155M}.  As a result, stars formed in this mass
range at redshifts $z>6$ can survive until the present day.  These
stars probe conditions during the formation of the Galaxy and can
likely reveal properties of the earliest, metal-free, so-called
Population~III stars.  Some of today's low-mass stars could themselves
be Population~III \citep{2015MNRAS.453.2771J} as recent simulations of
the formation of Population~III stars have increasingly favoured the
view that these stars could have had masses less than
1~M$_\odot$ \citep{2011ApJ...727..110C, 2011ApJ...737...75G,
2013ApJ...766..103D, 2014ApJ...781...60H, 2014ApJ...785...73S,
2014ApJ...792...32S, 2015MNRAS.448.1405M, 2016MNRAS.462.1307S}.

In the last few decades, significant effort has been made to find
metal-poor stars in the Galactic halo \citep{1981ApJ...248..606B,
1985AJ.....90.2089B, 1991AJ....101.1865R, 1992AJ....103.1987B,
1996AJ....112..668C, 2003RvMA...16..191C, 2007ApJ...670..774N,
2009AJ....137.4377Y, 2011Natur.477...67C, 2014Natur.506..463K} as well
as in the bulge
\citep{2016MNRAS.460..884H, 2016PASA...33...22N, 2015Natur.527..484H,
  2015ApJ...809..110C, 2015ApJ...807..171J}.  These surveys have
revealed about 400 stars with iron abundance
$[\mathrm{Fe}/\mathrm{H}]<-3$ and eight stars with
$[\mathrm{Fe}/\mathrm{H}]<-7.5$ to $-4.5$ \citep{2015ARA&A..53..631F},
based on high-resolution spectroscopic observations of candidate
extremely metal-poor stars.  Abundances of a variety of other elements
such as lithium, carbon, magnesium, and calcium have also been
measured in these stars.  None of these stars has the minimum
observable metal abundance, which is set by the capabilities of
spectrographs and currently corresponds to the metal abundance of a
cool red giant with $T_\mathrm{eff}=4500$~K and $\log g=1.5$ with
observed Ca~II~K line strength of
20~m\AA\ \citep{2015ARA&A..53..631F}.  Thus, no Population~III star
has yet been discovered.

Still, models of galaxy formation that account for metal production in
stars and supernova-driven metal enrichment of the interstellar medium
(ISM) can be used to infer constraints on early star formation,
including Population~III star formation, from the chemical properties
of observed metal-poor stars \citep{2001PhR...349..125B,
2004ARA&A..42...79B, 2005SSRv..116..625C, 2005SSRv..117..445G,
2006jebh.book..239R, 2006MNRAS.369..825S, 2007MNRAS.381..647S,
2009Natur.459...49B, 2009ApJ...696L..79K, 2010ApJ...717..542K,
2012MNRAS.423L..60S, 2013RPPh...76k2901B, 2013pss5.book...55F,
2014MNRAS.445.3039D, 2014ApJ...794..100M, 2015MNRAS.447.3892H,
2015MNRAS.452.2822S, 2016MNRAS.462.1307S, 2016arXiv161100759G,
2016ApJ...826....9I, 2017MNRAS.465..926D}.  For
example, \citet{2007MNRAS.381..647S} derived constraints on the
typical mass of Population~III stars $m_\mathrm{PopIII}$ and the
critical gas metallicity Z$_\mathrm{cr}$ below which Population~III
stars can form.  They argued that non-detection of a metal-free star
constrains Z$_\mathrm{cr}>0$ and $m_\mathrm{PopIII}>0.9$~M$_\odot$.
In their model, the observed metallicity distribution function (MDF)
of metal-poor stars in the Galactic halo was well-fit by
Z$_\mathrm{cr}=10^{-4}$~Z$_\odot$ and
$m_\mathrm{PopIII}=200$~M$_\odot$.  \citet{2012MNRAS.423L..60S} showed
that the elemental abundance pattern of the metal-poor star SDSS
J102915+172927, which has metallicity $Z=4.5\times 10^{-5} Z_\odot$
and mass less than 0.8 M$_\odot$, can be accounted for by the chemical
yields of core-collapse supernovae with metal-free progenitor stars
having masses 20 and 35 M$_\odot$.  By tracking the production of dust
in these supernovae, \citet{2012MNRAS.423L..60S} argued that the
existence of SDSS J102915+172927 implied that dust cooling played an
important role in the formation of the first low-mass Population~II
stars.  Recently, \citet{2015MNRAS.447.3892H} argued that if no
metal-free star is detected in a sample of $2\times 10^7$ halo stars
then $m_\mathrm{PopIII}>0.8$~M$_\odot$ at 99\% confidence level.

However, all of the above attempts at exploiting the metal abundance
properties of metal-poor stars assume that these stars preserve the
metal abundance properties of their natal star-forming clouds.  This
assumption can break down if stars are able to efficiently accrete
metals from the ISM as they orbit in the potential well of the Galaxy
throughout their lifetime.  Due to their large ages, low-mass
metal-poor stars can potentially interact with a large column of ISM
gas, depending on the stellar orbits.  During this time, the ISM
itself becomes increasingly enriched due to metal production from
other stars, as evidenced by the existence of an age-metallicity
relation in high-redshift galaxies \citep{2006ApJ...644..813E,
2008A&A...488..463M, 2012ApJ...755...89R}.  Accretion of metals from
this enriched ISM can pollute old metal-poor stars, and can result in
metal abundance patterns that can be quite different from the original
metal abundance patterns of these stars.  As a result, depending on
the magnitude of this effect, errors can be introduced in any
inference about high-redshift star formation drawn from the chemical
properties of old metal-poor stars.

Chemical enrichment of stars due to accretion from the ISM has been
studied over the last several decades \citep{1977ApJS...34..295T,
  1980ApJ...235..541A, 1981A&A....97..280Y, 1983MmSAI..54..321I,
  2009MNRAS.392L..50F, 2009ApJ...696L..79K, 2010ApJ...717..542K,
  2011MNRAS.413.1184J, 2014ApJ...784..153H, 2015ApJ...808L..47K, 2015MNRAS.453.2771J}.
Assuming that stars accrete only when they pass through the Galactic
disk and assuming a static Galactic disk with height 100~pc and
density $n\sim 5$~cm$^{-3}$, \citet{2009MNRAS.392L..50F} calculated
metal accretion for 474 stars using their kinematic measurements from
the Sloan Digital Sky Survey, and concluded that accretion is generally
negligible.  \citet{2009ApJ...696L..79K, 2010ApJ...717..542K,
  2015ApJ...808L..47K} considered metal accretion of metal-poor stars
in a semi-analytical galaxy formation model built on halo merger trees
obtained using the extended Press-Schechter formalism and concluded
that Population~III survivors can be enriched up to
$[\mathrm{Fe}/\mathrm{H}]\sim -5$ due to accretion from the ISM.
\citet{2011MNRAS.413.1184J} used a semi-analytical model of galaxy
formation built on average halo growth histories
\citep{2010MNRAS.406..896B} and assumptions about typical stellar
orbits to consider the accretion of metals on stars and its
suppression due to stellar winds. \citet{2015MNRAS.453.2771J}
performed a similar analysis to study the effect of radiation pressure
on metal accretion and argued that radiation pressure selectively
inhibits dust accretion.  This implies that stars enriched by
accretion from the ISM will show selective reduction in refractory
elements, which may explain the composition of at least some
carbon-enhanced metal-poor stars without neutron-capture element
overabundance (CEMP-no stars; \citealt{2005ARA&A..43..531B,
2016ApJ...833...20Y}, and references therein).  By comparing the
metallicities of low-mass stars with different spectral
types, \citet{2014ApJ...784..153H} found possible observational
evidence for metal accretion on the halo and thick-disk stars of the
Milky Way, and suggested that metal accretion could have occurred for
the stars in the satellite galaxies before they merged into the main
halo, since the velocity dispersions of these dwarf systems are much
smaller than the main progenitor of the Milky Way.

\begin{figure*}
  \begin{center}
    \includegraphics[width=\textwidth]{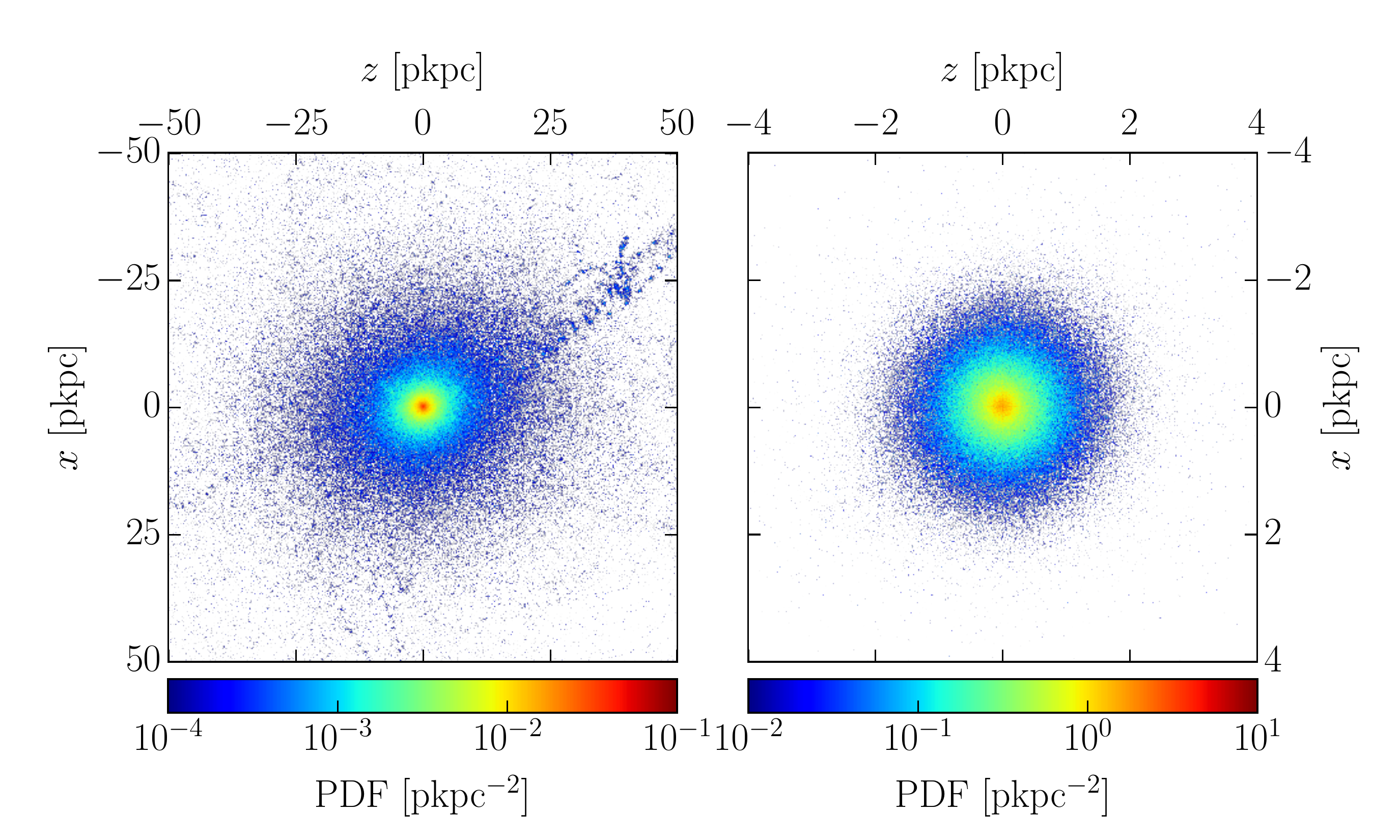}
  \end{center}
  \caption{Two-dimensional probability distribution functions derived
    from projected stellar orbits in Eris for halo (left panel) and
    bulge (right panel) stars.  Distances are in physical kiloparsecs
    (pkpc).  The projection shown here is over complete stellar orbits
    so that information from all redshifts ($z=0$--$12$) is combined.
    For comparison, the radial scale length disk of the Milky-Way
    analog in Eris is 2.5~kpc at $z=0$.  We find that both halo and
    bulge stars show significantly higher incidence in the central
    regions of the Galaxy.  As a result, stars in the halo as well as
    the bulge are likely to encounter dense gas, which can enhance
    accretion.  }
  \label{fig:orbits}
\end{figure*}

Nevertheless, these studies of metal accretion by stars simplify one
or both of two important aspects of the problem.  First, the dynamical
and chemical evolution of the ISM is often simplified, so that the
density evolution \citep{2013ApJ...773...43B} of gas in the Galactic
disk and halo as well as its chemical evolution
\citep{2011ApJ...729...16K, 2012ApJ...760...50S, 2013ApJ...772...93K,
  2013ApJ...765...89S, 2015ApJ...807..115S} are neglected.  In
reality, gas density not only evolves on average as $(1+z)^3$, but can
also briefly increase accretion rates on stars during episodes of
major galaxy mergers when the Galaxy has a disturbed morphology
\citep{2013ApJ...773...43B}.  Secondly, stellar orbits are also often
simplified.  As the Galaxy grows in mass via accretion and mergers,
its potential well deepens and virial velocity increases.  Major
mergers can have a strong effect on stellar orbits, which can increase
accretion rates for brief periods of time \citep{2013ApJ...773...43B}.
A large fraction of stars in the bulge and halo can be formed in
satellite progenitors rather than the main host halo
\citep{2015ApJ...799..184P}.  Also, before the Galaxy develops a
distinct halo, stars that are in the halo today can be on radial
orbits and can encounter denser-than-average gas, thus increasing the
metal accretion rates.  Our aim in this work is to achieve a more
realistic picture for both of these ingredients of stellar
metal-accretion models.

In this paper, we study the accretion of metals on halo and bulge
stars using the high-resolution cosmological zoom-in simulation
``Eris'' \citep{Guedes11}.  Eris creates a Milky Way analog at $z=0$
and represents a possible assembly history of the Galaxy.  Due to its
cosmological nature, Eris accounts for the cosmological inflow of
metal-poor gas as well as a plausible merger history of the Galaxy.
Due to its high resolution, Eris is able to evolve star particles that
represent the earliest stars on realistic orbits in an evolving
galactic potential.  Star formation and supernova implementations lead
to the chemical evolution of the ISM.  These properties of Eris make
it possible for us to investigate metal accretion on stars as they
interact with an ISM with realistic chemical and dynamical evolution.

We describe the Eris simulation in Section~\ref{sec:erissim}.
Section~\ref{sec:acc} details our metal-accretion model and our
results are presented in Section~\ref{sec:results}.  We end with a
summary of our main conclusions in Section~\ref{sec:conc}.

\section{The Eris Simulation}
\label{sec:erissim}

We use the high-resolution, zoom-in cosmological hydrodynamical
simulation ``Eris'' of a Milky Way Galaxy analog to investigate
accretion of metals onto old and metal-poor stars.  A detailed
description of the Eris simulation is provided by \citet{Guedes11}.
Here we briefly outline aspects relevant to this study.  The
simulation was performed with the parallel TreeSPH code
\textsc{Gasoline} \citep{Wadsley04} in a WMAP-3 cosmology
\citep{2007ApJS..170..377S}.  The halo has total mass $M_{\rm vir} =
7.9 \times 10^{11} \msun$ at $z = 0$, and was chosen to have a quiet
merger history with no major merger (of mass ratio greater than 1/10)
after $z$ = 3. The high-resolution region, centred around a Lagrangian
subregion of 1~Mpc on a side, contains 13 million dark matter
particles and an equal number of gas particles, for a final dark
matter and gas particle mass of $m_{\rm DM} = 9.8 \times 10^{4} \msun$
and $m_{\rm SPH} = 2.0 \times 10^{4} \msun$, respectively. The
gravitational softening length, $\epsilon_{\rm G}$, was fixed to 120
physical pc for all particle species from $z=9$ to the present, and
evolved as $1/(1+z)$ from $z=9$ to the starting redshift of $z=90$. A
total number of 400 snapshots were taken.  The time spacing between
two subsequent snapshots is about 30~Myr.

The simulation includes a uniform metagalactic UV background \citep{HM96},
Compton cooling, atomic cooling and metallicity-dependent radiative
cooling at T $<10^{4}$ K. Star formation is modelled by stochastically
forming ``star particles'' out of gas that is sufficiently cold ($T <
3 \times 10^{4}$ K) and reaches a threshold density of $n_{\rm SF}=5$
atoms cm$^{-3}$. The local star formation rate follows:
\begin{equation}
\frac{d\rho_*}{dt}=0.1 \frac{\rho_{\rm gas}}{t_{\rm dyn}} \propto \rho_{\rm
  gas}^{1.5},
\end{equation}
where $\rho_*$ and $\rho_{\rm gas}$ are the stellar and gas densities,
respectively, and $t_{\rm dyn}$ is the local dynamical time. Each star
particle has an initial mass $m_{\ast} = 6000 \ \msun $ and represents
a simple stellar population that follows a \citet{Kroupa93} initial
mass function (IMF). Note that this IMF is also applied for the first
generation of stars (Population III). In the simulation, new star
particles inherit the metallicity of their parent gas particles. Star
particles inject energy, mass and metals back into the ISM through
Type~Ia and Type~II SNe, and stellar winds \citep{Stinson06}. Eris's
high resolution enables the development of an inhomogeneous ISM, which
allows clustered star formation and strong cumulative feedback from
coeval supernova explosions. Large scale galactic winds are launched
as a consequence of stellar feedback, which transports a substantial
quantity of metals into the circumgalactic medium and enriches the
subsequent gas accretion \citep{2013ApJ...765...89S}. At $z = 0$, Eris
forms an extended, rotationally supported stellar disk with a small
bulge-to-disk ratio and a radial scale length of 2.5~kpc. The
structural properties, the mass budget in various components, and the
scaling relations in Eris are simultaneously consistent with
observations of the Galaxy \citep{Guedes11}.

The simulation follows \citet{Raiteri96} to model metal enrichment
from Type~Ia and Type~II SNe (\SNIa\ and \SNII).  Metals are
distributed to gas within the SPH smoothing kernel, which consists of
32 neighboring particles.  For \SNII, metals are released as the
main-sequence progenitors die, and iron and oxygen are produced
according to the following fits to the \citet{Woosley95} yields:
\begin{equation}                                                                
M_{\rm Fe} = 2.802 \times 10^{-4} \left({m_*\over
  \msun}\right)^{1.864}\,\msun,
\end{equation}                                                                  
and
\begin{equation}                                                                
M_{\rm O} = 4.586 \times 10^{-4} \left({m_*\over \msun}\right)^{2.721}\,\msun. 
\end{equation}
For \SNIa, each explosion produces 0.63~$\msun$ of iron and
0.13~$\msun$ of oxygen \citep{Thielemann86}.  The mass return from
stellar wind feedback is computed for stars with masses between 1 and
8~$\msun$.  For each stellar particle, the amount of stars lost as
supernovae during every timestep is computed following stellar
lifetimes by \citet{Raiteri96}, and the returned mass is calculated
using the initial-final mass relation of \citet{Weidemann87}.  The
mass is distributed to neighbouring gas particles in the same way as
the SNe feedback but with no energy injection.  The returned gas
inherits the metallicity of the star particle.  We adopt
the \citet{Asplund09} solar abundance scale for the elements other
than O and Fe that are not tracked in the simulation.

It is worth noting that the simulation used a traditional SPH
formalism where metals advect with the fluid perfectly without mixing
due to microscopic motions, and this may cause an artificially
inhomogeneous chemical distribution. Following
\citet{2015ApJ...807..115S}, to model the unresolved mixing we
post-process the simulation to assign each stellar particle with the
<{\it intrinsic} average metallicity of its 128 neighbouring gas
particles at the formation time. The corresponding mixing length is
generally around 50--120~pc at all redshifts. At our star formation
threshold density of 5 atoms cm$^{-3}$, this corresponds to a distance
that gas can cross within the free-fall time, assuming the typical
velocity dispersion of a molecular cloud. We have shown
in \citet{2015ApJ...807..115S} that, with this smoothing model, the
overall chemical evolution in Eris appears to reproduce the observed
[$\alpha$/Fe] as a function of metallicity in the Milky Way. When we
calculate the metal accretion rate (see Section~\ref{sec:acc} for
details), the gas properties are also smoothed over 128 neighbours.

\begin{figure*}
  \begin{center}
    \includegraphics[width=1.7\columnwidth]{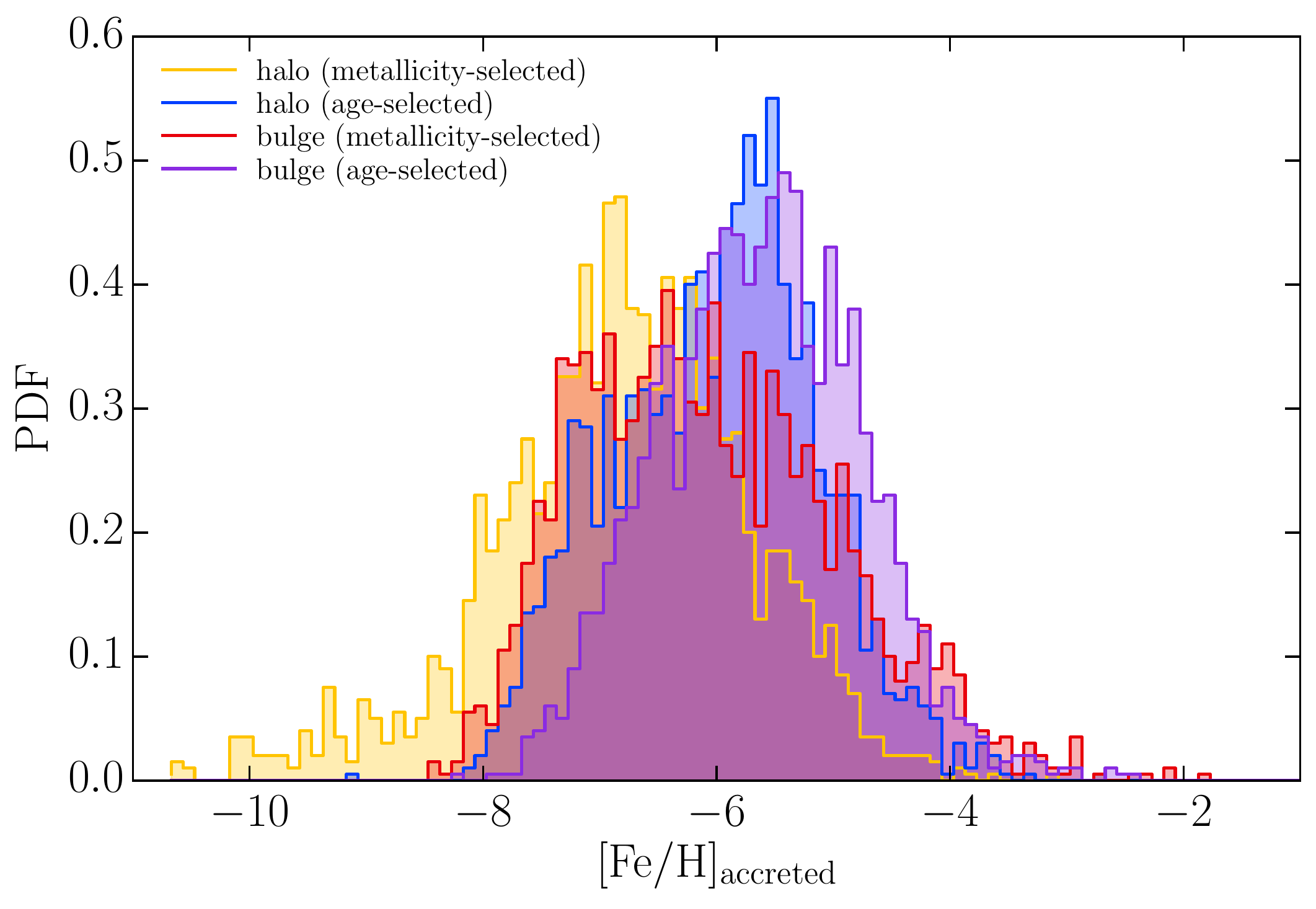}
  \end{center}
  \caption{Probability distribution functions (PDFs) of accreted iron
    abundance [Fe/H] onto stars in our sample.  Purple and red
    histograms show PDFs for bulge stars selected by their age and
    metallicity, respectively.  Blue and yellow histograms show PDFs,
    respectively, for age- and metallicity-selected halo stars.  We
    find that stars in our sample can be enriched up to
    $[\mathrm{Fe}/\mathrm{H}] \sim -6$ to $-5$ purely due to
    accretion.  The spread in these PDFs is due to the distribution of
    stellar orbits and gas metallicities in the Galaxy.  As a
    reference, the minimum observable iron abundance is about
    $[\mathrm{Fe}/\mathrm{H}]=-9.8$ \citep{2013pss5.book...55F}.}
  \label{fig:FeH_PDF}
\end{figure*}

\section{Metal-Accretion Model}
\label{sec:acc}

In Eris, stars are represented by collisionless star particles that
are created when gas particles satisfy the temperature and density
condition described in the previous section.  Each star particle has
an initial mass $m_{\ast} = 6000$~M$_\odot$, which reduces with time
due to stellar evolution.  While we cannot resolve individual stars
with mass of the order of $1$--$100$~M$_\odot$ due to unavoidable
limitations imposed by numerical resolution, star particles in Eris
still trace the underlying gravitational potential well.  Therefore,
we assume that these star particles provide a fair sample of the true
stellar orbits.  To calculate accretion of gas from the ISM onto
stars, we assume that stars follow the orbits of star particles in
Eris, and model accretion as a spherically symmetric Bondi-Hoyle flow,
given by
\citep{1939PCPS...35..405H, 1952MNRAS.112..195B, 1985MNRAS.217..367S}
\begin{equation}
  \dot{M} = \frac{2\pi(GM_\ast)^{2}\rho}{(v_\mathrm{rel}^{2} + c_{s}^{2})^{3/2}},
  \label{eqn:bh}
\end{equation}
where $M_{\ast}$ is the stellar mass, $\rho$ is the gas density,
$v_\mathrm{rel}$ is the relative velocity between the star and the
gas, and $c_{s}$ is the gas sound speed. For an ideal gas
$c_{s}=\sqrt{\gamma k_{B} T/\mu m_{H}}$, where $\gamma$ is the
adiabatic index, $k_{B}$ is the Boltzmann constant, $\mu$ is the mean
molecular weight and $m_{H}$ is the mass of the hydrogen atom. For
simplicity, we use a constant $\gamma = 5/3$ corresponding to an
adiabatic gas, and a constant $\mu = 1.2$, which corresponds to
primordial neutral gas.  In reality radiative cooling may reduce
$\gamma$ to a value between 5/3 and 1, photo-ionisation may decrease
$\mu$ by a factor of 2. These changes to $\gamma$ and $\mu$ may alter
the sound speed by at most 40\%.

We consider gas accretion onto four classes of stars, two of which are
defined by age and two by intrinsic metallicity: (1) stars formed in
the first 600~Myr of the simulation and classified as halo stars at
redshift $z=0$; (2) stars formed in the first 600~Myr of the
simulation and classified as bulge stars at redshift $z=0$; (3) stars
with intrinsic iron abundance as calculated in Eris
$[\mathrm{Fe}/\mathrm{H}]< -4$ and classified as halo stars at
redshift $z=0$; and (4) stars with intrinsic iron abundance
$[\mathrm{Fe}/\mathrm{H}]< -4$ and classified as bulge stars at
redshift $z=0$.  We thus consider accretion onto two classes of halo
stars and two classes of bulge stars.  Stars selected by metallicity
will in general have smaller ages than stars selected by our age
criterion.  Observed low-mass metal-poor stars lie at the intersection
of the two selections.  From a theoretical point of view, the two
selections may be expected to bracket extreme accretion scenarios, as
age-selected stars will accrete gas longer.  Note that the disk,
bulge, and halo of the Eris Galaxy analog at $z=0$ are determined by
kinematic decomposition as described by \citet{Guedes11}.  Of all the
Eris star particles falling in each group, we select a uniform random
sample of 2000 star particles to compute metal accretion.  Orbits of
these star particles are traced to the current epoch using 400
snapshots evenly spaced in time with $\Delta t \simeq$ 30 Myr.  For
each star particle, we compute the average gas sound speed $c_{s}$ and
metallicity over its 128 nearest neighbouring gas particles
\citep{2015ApJ...807..115S}, as described in
Section~\ref{sec:erissim}.  The relative velocity $v_\mathrm{rel}$ is
the velocity difference between the star particle and the
centre-of-mass velocity of the neighbouring gas particles.  The iron
or oxygen accretion rate is then given by multiplying the averaged
iron or oxygen mass fraction of the neighbouring gas by the mass
accretion rate from Equation~(\ref{eqn:bh}).  Figure~\ref{fig:orbits}
provides a visual impression of halo and bulge stars in Eris by
showing the two-dimensional probability distribution functions derived
from projected stellar orbits of a uniform random sample of halo and
bulge stars.  (Orbits of only metallicity-selected stars are shown.)
The projection is over complete orbits so that information from all
redshifts is combined.

We assume $M_*=0.8$~M$_\odot$ while evaluating the mass accretion rate
using Equation~(\ref{eqn:bh}).  This approximately corresponds to the
main-sequence turnoff mass in a 12~Gyr old stellar population.  The
maximum stellar mass for which a star in our age-limited sample can
survive up to the present day is 0.99~M$_\odot$.  The lifetime of a
star with mass 0.9~M$_\odot$ is 15~Gyr and that with mass
0.8~M$_\odot$ is 16.1~Gyr.  Stars selected by metallicity can have
much higher mass: metal-poor stars could form relatively recently in
pristine pockets of the ISM that are yet to be polluted by metals;
such stars can have higher mass and still survive to the present day.
Still, we find that 36\% of the bulge star particles selected by
metallicity form in the first 600~Myr of the simulation and 80\% form
in the first Gyr.  Of the halo stars selected by metallicity, 11\%
form in the first 600~Myr and 80\% form in the first 1.8 Gyr.  Metal
mixing in the ISM is efficient enough that most metallicity-selected
stars have large ages.  We assume that accreted metals are mixed in
surface convective zones with mass $3\times 10^{-3}$~M$_\odot$ for
$M_*=0.8$~M$_\odot$ \citep{1981A&A....97..280Y, 1995ApJ...444..175F}.
For a fixed convective zone mass fraction, the accretion rate scales
as $M_*^2$ and the relative metal abundance ratios [X/H] scale as
$M_*$.

\section{Results}
\label{sec:results}

With stellar orbits and gas metallicity predictions from Eris, and our
metal accretion model described in the previous section, we can now
consider metal enrichment of halo and bulge stars due to accretion.

\subsection{Metal accretion in Eris}

Figure~\ref{fig:FeH_PDF} shows probability distribution functions
(PDFs) of the relative abundance values [Fe/H] of accreted iron for
the four categories of stars defined in Section~\ref{sec:acc}.  The
four PDFs are quite similar, perhaps surprisingly so.  They peak
around $[\mathrm{Fe}/\mathrm{H}] \sim -6$ to $-5$, well within the
metallicity detection limit of modern surveys
\citep{2013pss5.book...55F} and comparable to the metallicities of
the most metal-poor stars known \citep{2011Natur.477...67C,
  2014Natur.506..463K}.  This indicates that accretion can have a
  significant impact on metal-poor main-sequence stars.  The four PDFs
  also show a considerable scatter---about three decades around their
  average values.  The remarkable agreement between the four
  distributions results from the fact that both halo and bulge stars
  encounter dense gas in the central regions of the Galaxy.  This is
  evident in Figure~\ref{fig:orbits}, which shows that both halo and
  bulge stars exhibit a significantly higher incidence in the central
  regions.  This is understandable, as a kinematically distinct halo
  and bulge do not exist in Eris at redshift $z>2$, and the last major
  merger of the Galaxy is at $z\sim 3$.  Additionally, the [Fe/H] PDFs
  of age- and metallicity-selected samples also agree quite well,
  because most metallicity-selected stars ($[\mathrm{Fe}/\mathrm{H}]<
  -4$) are quite old.  As mentioned above, 80\% of
  metallicity-selected halo stars were formed within the first 1.8~Gyr
  of the simulation, while the same fraction of metallicity-selected
  bulge stars were formed within the first 1~Gyr.  Thus, regardless of
  how they are selected at redshift $z=0$, halo as well as bulge stars
  in our analysis accrete to a level of $[\mathrm{Fe}/\mathrm{H}] \sim
  -6$ to $-5$ on average.

While the four distributions in Figure~\ref{fig:FeH_PDF} are
remarkably similar, there are small differences.  Age-selected bulge
stars have the highest median accretion
($[\mathrm{Fe}/\mathrm{H}]=-4.9$ with a 68\% spread of 1.7), followed
by age-selected halo stars ($[\mathrm{Fe}/\mathrm{H}]=-5.2$ with a
68\% spread of 1.8), and metallicity-selected bulge stars
($[\mathrm{Fe}/\mathrm{H}]=-5.4$ with a 68\% spread of 2.3).
Metallicity-selected halo stars have the lowest value,
$[\mathrm{Fe}/\mathrm{H}]=-6.0$ with a 68\% spread of 2.0.  The
age-selected sample of bulge stars has the highest accreted
metallicity, while the metallicity-selected sample of halo stars shows
the lowest accreted metallicity.  The PDFs of age-selected halo stars
and metallicity-selected bulge stars have intermediate values.  This
difference is still quite small compared to the overall spread of the
PDFs.  This ordering of the PDFs can also be understood from the small
differences in ages and orbital distribution of stars in the four
samples.  Age-selected stars in our analysis are marginally older than
the metallicity-selected stars, and therefore, on an average accrete
metals for a longer duration, resulting in higher [Fe/H].  Likewise,
bulge stars encounter marginally denser gas than halo stars, as seen
in Figure~\ref{fig:orbits}, which boosts the accretion rates onto
bulge stars.  Our finding that metal-poor bulge stars are generally
older than metal-poor halo stars is along the lines of previous
theoretical results \citep{2000fist.conf..327W, 2007ApJ...661...10B,
2010ApJ...708.1398T}.  This explains the ordering of the histograms
for metallicity-selected samples in Figure~\ref{fig:FeH_PDF}.

The result in Figure~\ref{fig:FeH_PDF} describes accretion onto
0.8~M$_\odot$ main-sequence stars, but it is possible to understand
how it will change for different stellar masses.  For a fixed
convective-zone mass fraction, the accreted metal abundance [Fe/H]
scales with $M_*$.  For smaller stellar masses, the enrichment levels
will drop due to the reduced accretion rate and enhanced convective
layer mass.  For a 0.6~M$_\odot$ main-sequence star, the convective
zone mass increases to $\sim
10^{-2}$~M$_\odot$ \citep{1981A&A....97..280Y} while the accretion
rate drops by a factor of 1.8.  As a result, the histograms in
Figure~\ref{fig:FeH_PDF} will shift towards lower values of [Fe/H] by
about two decades.  Old stars with masses greater than 0.8~M$_\odot$
will likely be on the giant branch today.  Although accretion rates of
these stars remain quite high, the resultant [Fe/H] ratios drop
because of larger convective zone masses, which are of the order of
$10^{-1}$~M$_\odot$ for giants \citep{1981A&A....97..280Y}.  In this
case the histograms in Figure~\ref{fig:FeH_PDF} will shift towards
lower values of [Fe/H] by almost three decades.
Figure~\ref{fig:FeH_PDF} thus presents as extreme case.  Negative
feedback from processes such as stellar winds and radiation feedback
may also reduce accretion \citep{2015MNRAS.453.2771J,
2011MNRAS.413.1184J}.  Our results from Figure~\ref{fig:FeH_PDF}
predict marginally higher accreted metallicity than the model of
\citet{2009MNRAS.392L..50F}.  Our results are closer to the extreme
model of \citet{2009MNRAS.392L..50F}, in which it is assumed that the
stars pass once through a dense cloud with density $n\sim
10^3$~cm$^{-3}$.  This difference is mainly because the gas density
increases quite rapidly towards high redshift, while
\citet{2009MNRAS.392L..50F} only consider accretion from a static gas
disk with height 100~pc and density $n\sim 5$~cm$^{-3}$.  The median
accreted [Fe/H] in their fiducial model is about $-7$.  This increases
to $-5$ in their extreme model.  The latter value is closer to the
median values of all four samples of stars in
Figure~\ref{fig:FeH_PDF}.  Our median values are in agreement with
predictions by
\citet{2015ApJ...808L..47K} obtained from a semi-analytical galaxy
formation model built on halo merger trees.

\begin{figure}
  \includegraphics[width=\columnwidth]{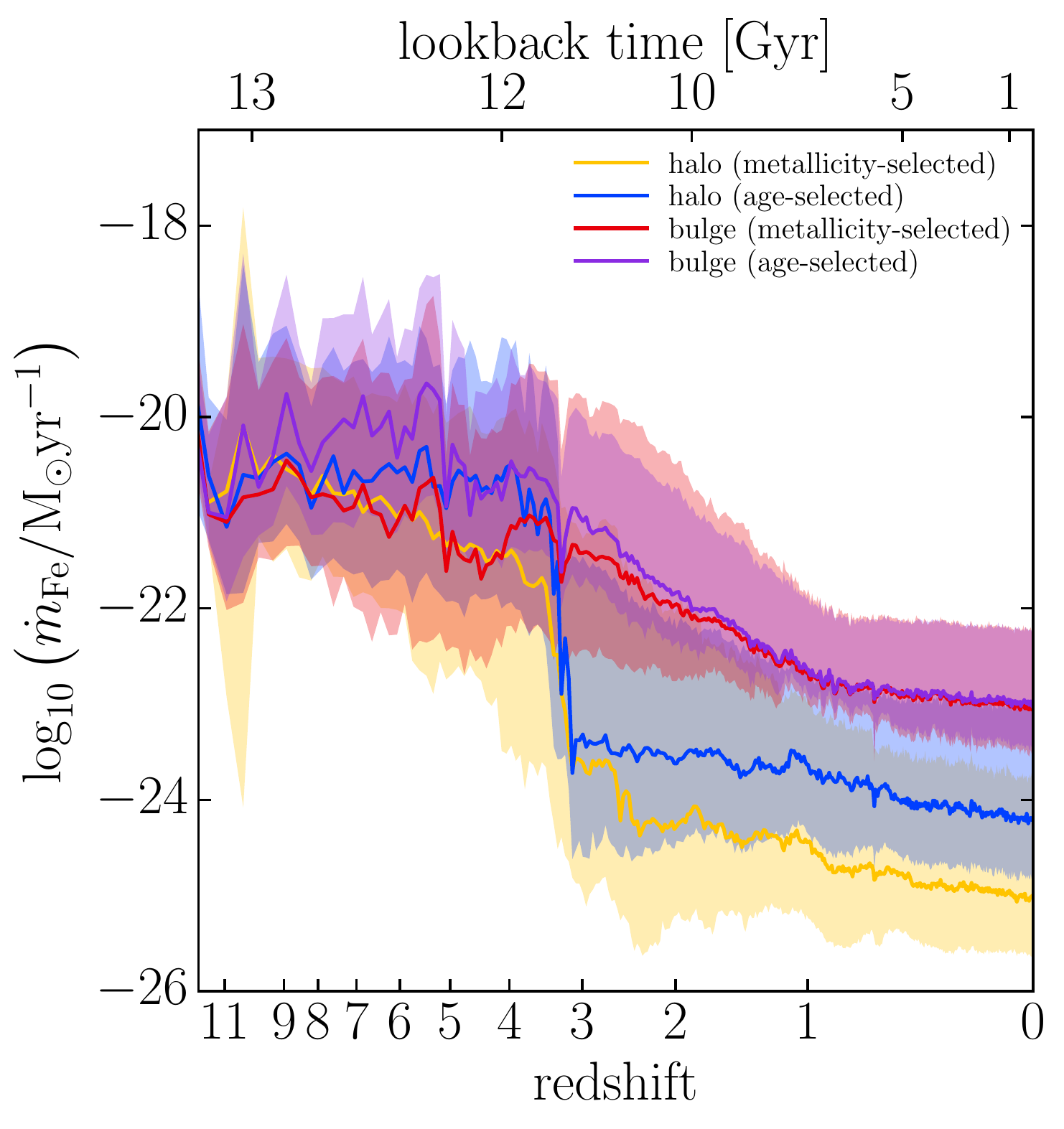}
  \caption{Evolution of the metal-accretion rate in our four samples
    of stars.  Solid curves show the median values and shaded regions
    show 68\% scatter around the median.  Blue and yellow curves and
    shaded regions describe properties of halo stars selected by age
    and metallicity, respectively.  Bulge stars selected by age and
    metallicity are described, respectively, by purple and red curves.
    The sudden drop in the accretion rates onto halo stars at $z\sim
    3.1$ is associated with the last major merger of the galaxy.}
  \label{fig:accrate}
\end{figure}

\begin{figure*}
  \includegraphics[width=\textwidth]{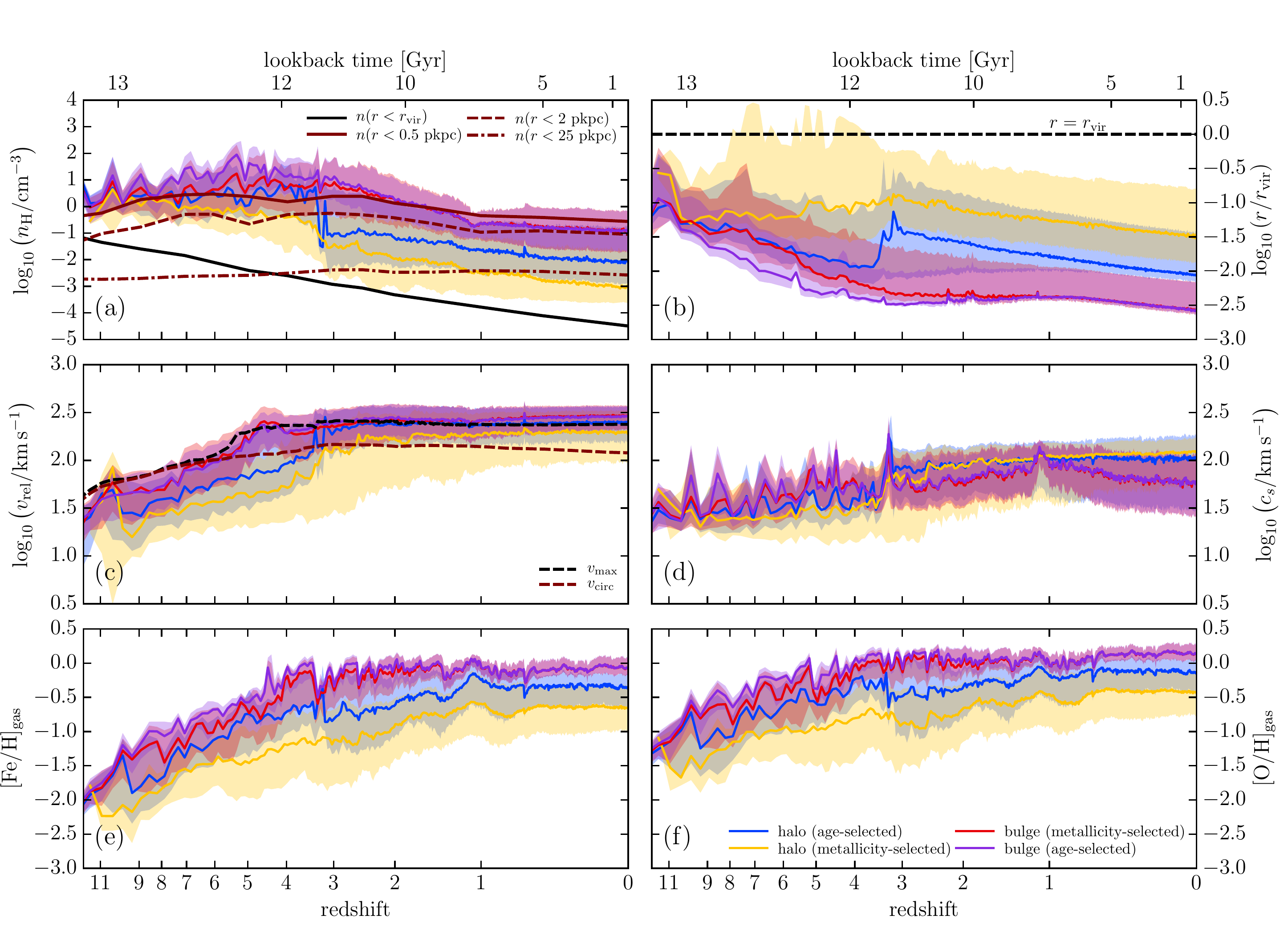}
  \caption{Evolution of various quantities in Eris: (a) gas density;
    (b) radial distance of stars from centres of mass of respective
    host halos; (c) stellar relative velocity with respect to the gas;
    (d) gas sound speed; (e) gas iron abundance [Fe/H]; (e) gas oxygen
    abundance [O/H].  Gas densities, sound speed, and metal abundances
    refer to gas along stellar orbits.  In each panel, solid curves
    show median quantities and shaded regions show 68\% scatter around
    the median.  Blue and yellow curves and shaded regions describe
    properties of halo stars selected by age and metallicity,
    respectively.  Bulge stars selected by age and metallicity are
    described, respectively, by purple and red curves.  In panel (a),
    the black curve shows the average gas density within the virial
    radius.  Brown dot-dashed, dashed, and solid curves in this panel
    show evolution of density averaged within 25, 2, and 0.5 pkpc from
    the centres of mass of progenitor haloes of the Eris Milky Way.
    In panel (c), the brown dashed curve shows the evolution of the
    circular velocity $v_\mathrm{circ}$ at the virial radius.  The
    black dashed curve in this panel shows the evolution of the
    maximum rotational velocity $v_\mathrm{max}$.  Densities and
    velocities are in physical units.  }
  \label{fig:evolution}
\end{figure*}

\subsection{Understanding metal-accretion rates}

In order to understand the stellar metal-accretion levels in our
model, Figures~\ref{fig:accrate} and \ref{fig:evolution} show the
evolution of various relevant quantities in Eris for the four samples
of stars considered.  Figure~\ref{fig:accrate} shows the evolution of
the metal-accretion rate of iron.  Yellow and blue curves correspond
the metallicity- and age-selected samples, respectively, of halo
stars.  Red and purple curves correspond to bulge stars.  Solid curves
show the evolution of the median value; shaded regions in
corresponding colours show ranges of 68\% scatter around the medians.
We see that the average metal accretion rate increases as we go
towards higher redshifts.  All four categories of stars have very
similar accretion rates at redshifts $z>3$.  It is only at lower
redshifts ($z<3$), after the last major merger of the Eris Milky Way,
that halo and bulge stars kinematically separate with halo stars
occupying regions with low gas density and therefore low accretion
rates.  The average metal accretion rate at high redshifts is about
$10^{-20}$ M$_\odot$yr$^{-1}$.  By $z=0$, this drops by two orders of
magnitude for bulge stars and by four orders of magnitude for halo
stars.  Note that the difference between the metal accretion rates of
age- and metallicity-selected stars is relatively small.  This is
simply because our metallicity-selected samples are also predominantly
composed of old stars, as explained above.

The relatively high values of accretion rates at high redshift are
driven by the gas density evolution.  This is evident in panel~(a) of
Figure~\ref{fig:evolution}, which shows the evolution of the average
physical density of gas along the stellar orbits in each of our four
samples.  The difference between the four curves is along the lines of
what we expect---metallicity-selected halo stars encounter the lowest
gas density, while the age-selected bulge stars encounter the highest
gas density.  This difference explains the trends in the accretion
rates shown in Figure~\ref{fig:accrate} as well as in the histograms
of accreted [Fe/H] in Figure~\ref{fig:FeH_PDF}.  The black curve in
panel~(a) shows the evolution of the average density within the virial
radius of the halo.  The virial density evolves at $(1+z)^3$.  A
comparison between the virial density curve and the median gas density
curves shows that the density of gas encountered by stars in our
samples grows roughly as $(1+z)^3$ from redshift $z=0$ to $6$, and
then drops because the halo turns around.  Importantly, the stars in
our samples encounter a significantly higher gas density than the
virial density as both halo and bulge stars are centrally concentrated
compared to the virial radius, as in Figure~\ref{fig:orbits}.  This
can be understood further with help from the brown dot-dashed, dashed,
and solid curves in panel~(a) of Figure~\ref{fig:evolution}.  These
curves show the average gas density in Eris within 25, 2, and 0.5
physical kpc (pkpc), respectively.  At high redshifts, $z>3$, almost
all stars in all four of our samples are interacting with gas within
in the inner 2 pkpc of the progenitor haloes of the Milky Way.  This
gas density is considerably higher than the gas density assumed in
previous works \citep[cf.][]{2009MNRAS.392L..50F,
2011MNRAS.413.1184J}, which results in the high metal accretion seen
in Figure~\ref{fig:FeH_PDF}.  This picture is reinforced by panel~(b)
of Figure~\ref{fig:evolution}, which shows the evolution of the radial
distribution of the stars in our four samples.  The solid curves in
this panel show the radial distance of stars from centres of mass of
respective host halos with the scatter represented by the shaded
areas.  All four samples of stars are quite centrally concentrated.
The halo stars generally have larger radial separations than the bulge
stars, which explains the density trends seen in panel~(a).

In panel~(c), Figure~\ref{fig:evolution} shows the evolution of the
relative velocity of stars with respect to ambient gas.  The velocity
decreases with increasing redshift due to the decreasing halo mass.
This also increases the accretion rate.  While accretion rates at high
redshift are enhanced by increased gas density and reduced velocities,
the gas metallicity evolution restricts the enhancement.  The brown
dashed curve in this panel shows the evolution of the circular
velocity $v_\mathrm{circ}$ at the virial radius, while the black
dashed curve shows the evolution of the maximum rotational velocity
$v_\mathrm{max}$ in the main progenitor halo of the Eris galaxy.
Stars in all four samples tend to have lower velocity than
$v_\mathrm{max}$, as the rotation curve drops at small radii.  The
stellar velocities can also be compared with the evolution of the gas
sound speed, which is shown in panel~(d).  The sound speed decreases
towards higher redshift with reduced gas temperature.  Panels~(e) and
(f) of Figure~\ref{fig:evolution} show the evolution of the iron and
oxygen abundances of the gas along orbits of stars in our four
samples.  There is a metallicity gradient in the ISM with most
metal-enriched gas present near the center of the halo.  This
metallicity gradient is also seen in other simulations of the Galaxy's
assembly \citep[e.g.,][]{2013MNRAS.436..625S}.  As a result, bulge
stars accrete relatively more metal-rich gas compared to halo stars.
The metallicity of gas around bulge stars evolves from
$[\mathrm{Fe}/\mathrm{H}]=-1.5$ at $z\sim 10$ to
$[\mathrm{Fe}/\mathrm{H}]=0.0$ at $z=0$.  The oxygen abundance evolves
from $[\mathrm{O}/\mathrm{H}]=-1.0$ to $[\mathrm{O}/\mathrm{H}]=0.1$
in this redshift range.  The metallicity of gas around halo stars is
lower by about 0.7.  The resultant $\alpha$-enhancement of ISM gas at
high redshift by about 0.5 relative to low redshifts is seen in
panel~(e) of Figure~\ref{fig:evolution}, which shows the evolution of
$[\mathrm{O}/\mathrm{Fe}]$.

\begin{figure}
  \begin{center}
    \includegraphics[width=\columnwidth]{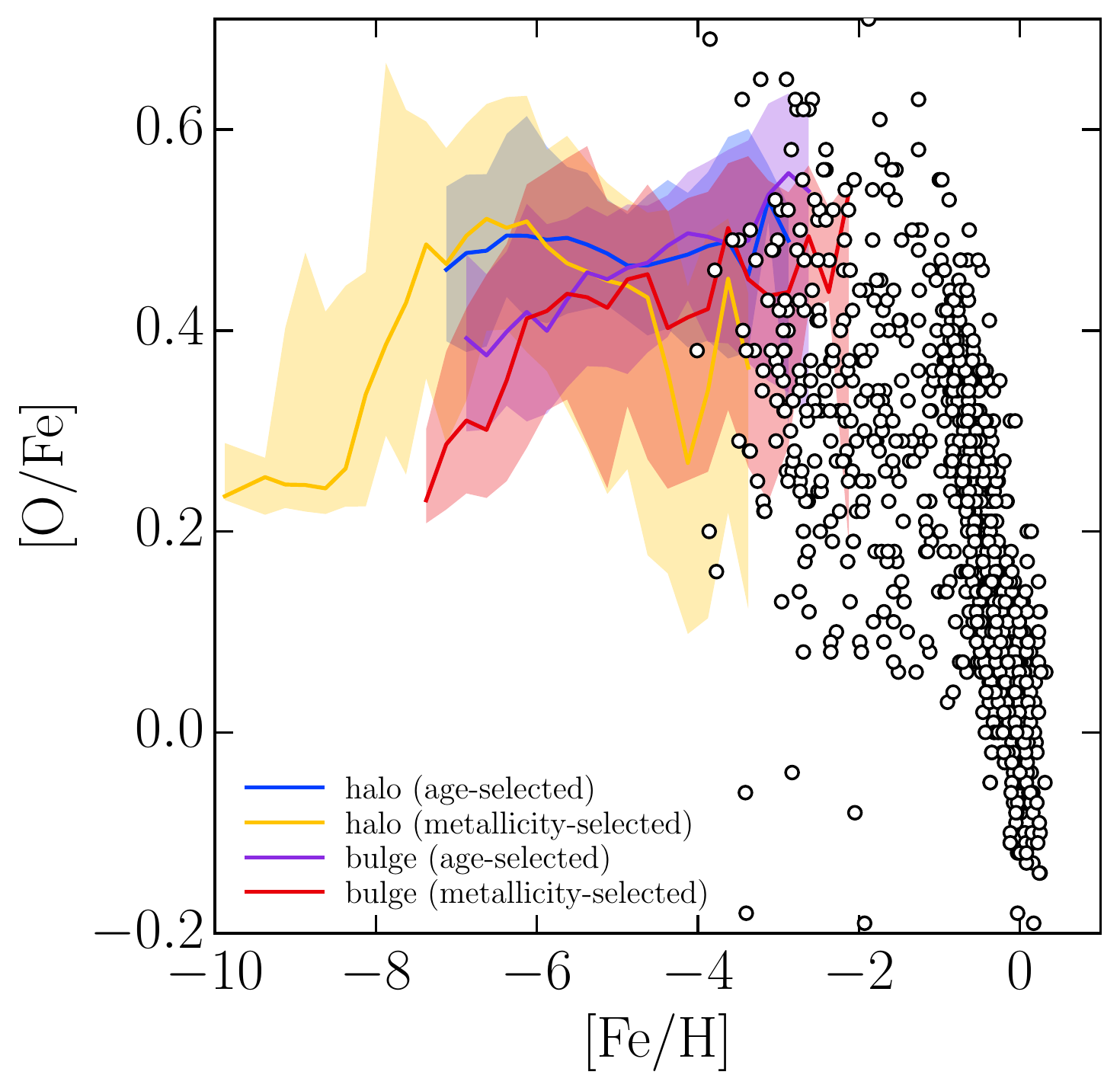}
  \end{center}
  \caption{Median $\alpha$-enhancement [O/Fe] of accreted gas by stars
    in the four samples considered in this paper (solid curves) and
    68\% scatter around the median (shaded regions).  Blue and yellow
    curves and shaded regions describe properties of halo stars
    selected by age and metallicity, respectively.  Bulge stars
    selected by age and metallicity are described, respectively, by
    purple and red curves.  Open circles show $[\alpha/\mathrm{Fe}]$
    measurements for halo and disk stars from the compilation by
    \citet{2015ApJ...807..115S}.  There is a good agreement between
    the $[\alpha/\mathrm{Fe}]$ for iron-poor stars and the accreted
    gas in our model.}
  \label{fig:ofe}
\end{figure}

Figure~\ref{fig:ofe} shows the resultant [O/Fe] ratio at $z=0$ due to
accretion in the four classes of stars considered here, as a function
of the accreted [Fe/H].  This $\alpha$-enhancement is shown in
comparison with observed [$\alpha$/Fe] ratios from disk and halo stars
in the Galaxy \citep{2000AJ....120.1841F, 2003MNRAS.340..304R,
2006MNRAS.367.1329R, 2004A&A...416.1117C, 2004ApJ...617.1091S,
2004AJ....128.1177V, 2005A&A...439..129B, 2013A&A...552A.128M,
2014AJ....147..136R}.  The accreted [O/Fe] ratios are in agreement
with the [O/Fe] ratios seen in the most metal-poor stars (see
also \citealt{2006ApJ...648..383P}).  At iron abundances of
$[\mathrm{Fe}/\mathrm{H}]>-1$ the observed [O/Fe] drops sharply to
less than $\sim 0.2$, signalling the effect of SN~Ia
\citep{1979ApJ...229.1046T}.  The accreted $\alpha$-enhancements are
much higher.  The reason is clear from
Figure~\ref{fig:accrate}---stars in all four of our samples accrete
most of their metals at high redshifts before SN~Ia start playing an
important role.  Thus, most accreted gas is $\alpha$-enhanced (cf.\
panels~(e) and (f) of Figure~\ref{fig:evolution}).  This also tells us
that measurements of [$\alpha$/Fe] will not help in separating stars
with intrinsic metal enrichment from stars that have accreted their
metals.  Perhaps the only way to disentangle these stars would be to
observe a peculiar enrichment pattern that is unlikely to exist in the
low-redshift ISM, such as the enhanced [Si/C] ratio or the large
contrast between the abundances of odd and even element pairs that is
predicted for pair-instability
supernovae \citep{2002ApJ...567..532H,2014Sci...345..912A}.

\subsection{Effect of accretion on the MDF}

\begin{figure}
  \includegraphics[width=\columnwidth]{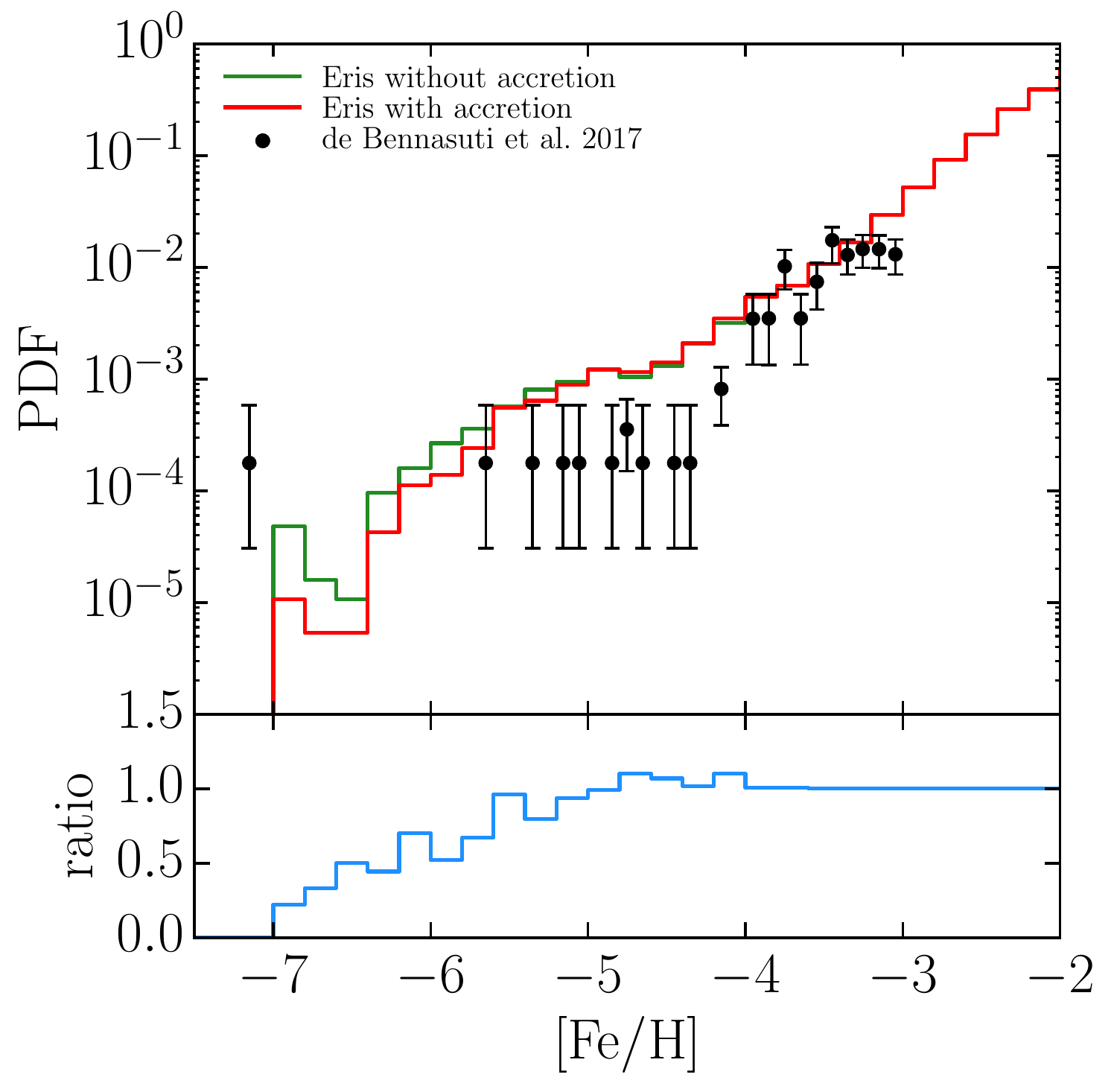}
  \caption{The metallicity distribution function (MDF) of halo stars
    in Eris simulation with (red curve) and without (green curve)
    accounting for metal accretion by stars from the ISM.  The data
    points from the compilation by \citet{2017MNRAS.465..926D}.  The
    blue curve in the lower panel shows the ratio of the two MDFs,
    illustrating the truncation of the low-metallicity tail of the
    MDF due to accretion.}
\label{fig:mdf_corrected}
\end{figure}

Most models of Galactic chemical enrichment---including our base Eris
simulation---assume that enrichment of stars due to accretion of gas
from the ISM is negligible.  In Figure~\ref{fig:mdf_corrected} we
consider the error incurred due to this assumption.  Black symbols in
this figure show the measured metallicity distribution function (MDF)
as compiled by \citet{2017MNRAS.465..926D}, shown here as a
probability distribution function.  The green histogram shows the MDF
of halo stars from Eris without any correction due to accretion.  This
histogram refers to the intrinsic metallicity of halo stars in Eris.
The red histogram in Figure~\ref{fig:mdf_corrected} shows how the
black histogram is modified when accretion is taken into account.  In
each [Fe/H] bin, stars are added from lower metallicity bins by
gaining excess [Fe/H] due to accretion, in proportion to the PDF
derived in Figure~\ref{fig:FeH_PDF}.  Similarly, each [Fe/H] bin loses
stars to higher metallicity bins due to accretion.  Only the
metallicity-selected sample of halo stars (which has stars that have
$[\mathrm{Fe}/\mathrm{H}]<-4$ at $z=0$) is used for this purpose.  The
predicted MDFs agree quite well with observations.

As described in previous sections, enrichment due to accretion is
small ($[\mathrm{Fe}/\mathrm{H}]\sim -6$ to $-5$ on average).
Therefore, accretion affects only the metal-poor tail of the MDF; the
MDF at high metallicities remains unmodified.  However, the effect on
the MDF at $[\mathrm{Fe}/\mathrm{H}]<-5$ is considerable, suggesting
that a large fraction of metal-poor stars in this regime could be
potentially enriched due to accretion.  This is emphasised by the
bottom panel of Figure~\ref{fig:mdf_corrected}, which shows the ratios
of the uncorrected and corrected MDFs from the top panel.  Accretion
truncates the low-metallicity end of the MDFs to about
$[\mathrm{Fe}/\mathrm{H}]=-7$. Metal abundance measurements of these
stars would then contain little or no information about the chemistry
of their natal environments, thus affecting any inference of the
Population III IMF from these
stars \citep[e.g.,][]{2007MNRAS.381..647S, 2012MNRAS.423L..60S,
2015MNRAS.447.3892H, 2017MNRAS.465..926D}. It should also be noted
that accretion may change the relative abundances of elements by a
greater degree than the absolute abundance of a single element.
\citet{2015MNRAS.453.2771J} has proposed that if stars do indeed
accrete metals from the ISM, then radiation pressure would selectively
inhibit accretion of dust grains.  As a result, if most metal
enrichment in these stars is due to accretion, then refractory
elements such as iron would appear to be selectively depleted.  This
pattern of metal enrichment is tantalizingly similar to that seen in
carbon-enhanced metal poor (CEMP) stars \citep[][and references
therein]{2005ARA&A..43..531B,2016ApJ...833...20Y}. Seven of the eight
stars with $[\mathrm{Fe}/\mathrm{H}]<-4.5$ are carbon-enhanced.  The
one star with low carbon abundance \citep{2011Natur.477...67C} could
still be a CEMP star with a carbon abundance that is yet to be
detected, possibly because of the limited spectral S/N and warm
temperature.\footnote{This is the case, for instance, for the Group~II
CEMP-no stars G64-12 and G64-37, which were not recognised as CEMP
until extremely high S/N spectra were
obtained \citep{2016ApJ...829L..24P}.}  The argument
of \citet{2015MNRAS.453.2771J} would then suggest that the known
metal-poor stars have been enriched mostly by accretion of metals from
the ISM.  This is in good agreement with what we find in
Figure~\ref{fig:mdf_corrected}.

\section{Conclusions}
\label{sec:conc}

We have studied the chemical enrichment of stars due to accretion of
metal-enriched gas from the ISM during the Milky Way's development,
using the Eris zoom-in cosmological simulation of assembly of a Milky
Way analog.  We considered metal-poor and old stars in both the
Galactic halo and bulge and make use of the stellar orbits and gas
density and metallicity distributions predicted by Eris.

Our main conclusion is that both halo and bulge stars accrete metals
from the ISM and, in extreme cases, can reach enrichment levels of
$[\mathrm{Fe}/\mathrm{H}]\sim -2$ due to accretion alone.  The median
enrichment level is about $[\mathrm{Fe}/\mathrm{H}] \sim -6$ to $-5$
for age- as well as metallicity-selected stars in both the halo and
bulge.  There is little difference in the accreted enrichment levels
of the four classes of stars, although metallicity-selected halo stars
tend to accrete the least amount of metals.  Metal-poor bulge stars
are generally older than metal-poor halo stars.  As a result, bulge
stars show a marginally higher degree of effect of accretion.  Halo
and bulge stars accrete metals at the rate of about $10^{-24}$
M$_\odot$yr$^{-1}$ and $10^{-22}$ M$_\odot$yr$^{-1}$, respectively, at
redshifts $z\lesssim 3$, but this accretion rate increases a
hundred-fold to about $10^{-20}$ M$_\odot$yr$^{-1}$ at higher
redshifts due to increased gas density.  The drop in accretion rates
with time is mainly because of decreasing gas density.  Bulge and halo
stars accrete similar amounts of metals at high redshifts as
kinematically distinct bulge and halo are not yet developed at these
redshifts and both sets of stars encounter similar metal distribution
in the ISM on average.

Because accretion mostly takes place at high redshifts, it is
$\alpha$-enriched to $[\alpha/\mathrm{Fe}]\sim 0.5$.  This is
comparable to the $\alpha$-enrichment levels seen in the most
metal-poor stars.  It also means that $\alpha$-enrichment does not act
as a discriminant between intrinsically and extrinsically enriched
stars.  Accretive metal enrichment is sufficient to change the
predicted MDF of halo stars at $[\mathrm{Fe}/\mathrm{H}]<-5$.  This
suggests that attempts to infer the natal chemical environment of the
most metal-poor stars from their observed enrichment today can be
hindered due to metal accretion.  Our analysis assumes a stellar mass
of 0.8~M$_\odot$.  Accreted metal enrichment level drops rapidly at
smaller stellar masses due to decreasing accretion rates and increased
convective zone masses.  This, faint dwarfs will be more likely to
probe the first generation of stars.  Peculiar enrichment patterns
such as those predicted to arise from pair-instability supernovae
could help in disentangling natal and accreted metal content of stars
\citep{2014Sci...345..912A}.  Most stars are rich in oxygen and
carbon with $[\mathrm{O}/\mathrm{H}]<-3.2$ and
$[\mathrm{C}/\mathrm{H}]<-3.5$.  The accreted gas in our model only
has $[\mathrm{O}/\mathrm{Fe}]=0.5$.  Therefore, although the iron
abundance is significantly enhanced by accretion the abundance in
elements such as C and O should still remain unchanged and could be
useful in probing Population~III.  Also, our findings suggest that
lowest-metallicity Damped Lyman-$\alpha$
Systems \citep{2006A&A...451...19E, 2012MNRAS.421L..29S,
2012MNRAS.425..347C, 2013ApJ...772...93K} would provide an important
cross-check for Population~III yields inferred from metal-poor stars.

\section*{Acknowledgements}

We thank the referee, Timothy Beers, for his positive comments and
constructive suggestions, and also acknowledge helpful discussions
with Andrew Casey, Ryan Cooke, Denis Erkal, Gerry Gilmore, Martin
Haehnelt, Kohei Hattori, Joseph Hennawi, Sergey Koposov, Thomas
Masseron, Max Pettini, Enrico Ramirez-Ruiz, Alberto Rorai, and
Stefania Salvadori.  S.S. gratefully acknowledges support by Science
and Technology Facilities Council and by ERC Starting Grant 638707
`Black holes and their host galaxies: coevolution across cosmic time'.
G.K. and S.S. acknowledge support from ERC Advanced Grant 320596 `The
Emergence of Structure During the Epoch of Reionization'. Support for
this work was provided by NASA through grant HST-AR-13904.001-A
(P.M.). P.M. also acknowledges a NASA contract supporting the
WFIRST-EXPO Science Investigation Team (15-WFIRST15-0004),
administered by GSFC, and thanks the Pr\'{e}fecture of the
Ile-de-France Region for the award of a Blaise Pascal International
Research Chair, managed by the Fondation de l'Ecole Normale
Sup\'{e}rieure.  This work used the DiRAC/Darwin Supercomputer hosted
by the University of Cambridge High Performance Computing Service
(http://www.hpc.cam.ac.uk/), provided by Dell Inc.  using Strategic
Research Infrastructure Funding from the Higher Education Funding
Council for England and funding from the Science and Technology
Facilities Council.

\bibliographystyle{mnras}
\bibliography{refs}

\bsp
\label{lastpage}
\end{document}